\documentclass{llncs}
\usepackage{makeidx}  
\usepackage{algorithm}
\usepackage[noend]{algpseudocode}
\usepackage{graphicx}
\usepackage{hyperref}
\graphicspath{ {images/} }
\usepackage{booktabs}
\usepackage[spanish]{babel}
\usepackage[utf8]{inputenc}
\begin{document}
\frontmatter          
\pagestyle{headings}  
\addtocmark{Hamiltonian Mechanics} 

\mainmatter              
\title{Un caso de big data punta a punta: análisis de datos de transporte y su uso en el negocio.}
\titlerunning{Análisis de datos de transporte}  
%

\author{Camilo Melani,  Juan V. Echagüe, Joaquín Torre Zaffaroni, Daniel Yankelevich}
\institute{Pragma Consultores, San Martín 550 | (C1004AAL) Buenos Aires - Argentina
\email{dyankelevich@pragmaconsultores.com}}
%
%
%

\pagestyle{empty}
\maketitle              


%
\section{Introducción}
En este artículo se presentan los resultados de un proyecto de análisis de datos de una empresa de transporte, que involucró la recolección, preparación, visualización, transformación y análisis de 3 años de datos de viajes de colectivos, incluyendo boletos y posicionamiento geográfico. Este caso cubre el proyecto de punta a punta, incluyendo la incorporación de los resultados en el proceso de negocio.

Nosotros sostenemos que una característica clave de los proyectos de big data debe encontrarse en el proceso que se lleva a cabo y que inicia con la captura de grandes cantidades de datos, pasando por el procesamiento (que en muchos casos requiere una infraestructura especial o particular, con más de una computadora, en modo distribuido) hasta el análisis y el aprovechamiento de la información en el negocio. En nuestro punto de vista, este último paso (la inserción de la información en la toma de decisiones del negocio) es tan importante como el uso de bases NoSQL o Hadoop o procesar varios terabytes.

\section{Datos}
\label{sec:analisisrealizados}

SUBE es una tarjeta prepaga emitida por el Gobierno Nacional argentino para facilitar la movilidad en el área metropolitana. Puede usarse en los medios de transporte públicos en la región Metropolitana de Buenos Aires y el interior del país. La red de uso está compuesta por 11.000 colectivos, 5 líneas de subtes y las líneas ferroviarias metropolitanas, y diariamente vende 12 MM de boletos de transporte.

\section{Infraestructura}
Para realizar el procesamiento de los datos es necesario contar con capacidad de almacenamiento, acceso a la información y poder de cálculo adecuados. Para este caso utilizamos infraestructura propia con tecnología Apache HDFS \cite{ApacheHDFS}, Hive \cite{ApacheHive}, R \cite{R} y Hadoop \cite{Hadoop} sobre una estructura de 6 nodos.

\section{Limpieza, comprensión y análisis de datos}
La preparación de datos es una parte importante en un proyecto de big data \cite{HernandezDataCleansing}, de hecho en muchos casos el “data cleansing” y preparación inicial toma más tiempo que el análisis. En este proyecto, la preparación de datos incluyó identificar y subsanar varias limitaciones de los datos, por ejemplo, los relojes de los lectores del sistema SUBE y los GPS no están sincronizados. Asimismo, el trabajo se realizó sobre datos anónimos lo que requirió trabajo adicional.

\begin{figure}
  \centering
    \includegraphics[width=9cm,height=8cm]{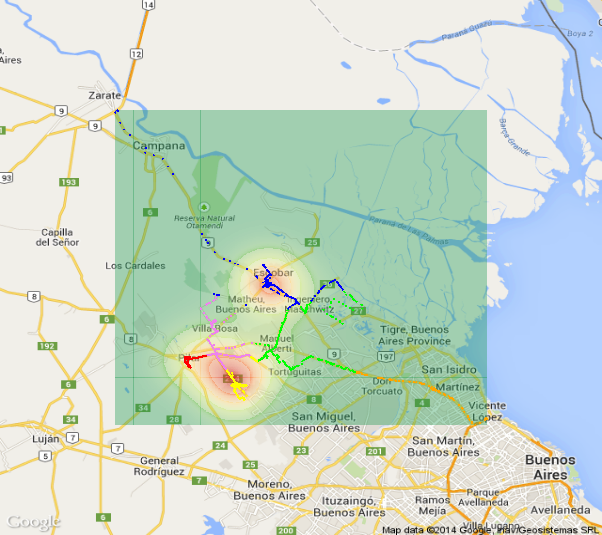}
  \caption{Clustering de la posición de la venta de boletos. Heatmap de concentracion de puntos.}
  \label{fig:clustering}
\end{figure}

Las tareas de análisis incluyeron la elaboración de histogramas, gráficos de series temporales, heatmaps en varias variables, generación de imágenes geo localizadas de la concentración de venta de boletos, identificar los trayectos de mayor demanda, relacionar los pasajeros frecuentes con el tiempo entre trayectos y generación de grafos.
Gran parte del análisis se focalizó en lo que identificamos como casos o preguntas del negocio: qué era lo que el negocio consideraba interesante para conocer y a qué le otorgaba valor, identificar el comportamiento de los clientes y características que permitieran su segmentación. El hecho de contar con toda la serie histórica desde que se implementó la tarjeta SUBE en esta empresa, nos permitió observar con sumo detenimiento la curva de adopción del sistema y el comportamiento de reemplazo del modelo anterior. Este mecanismo permite analizar y establecer patrones sobre el proceso de adopción de políticas públicas. Encaramos un estudio multiescala sobre la densidad de venta de boletos en diferentes horarios (ver Fig. \ref{fig:clustering}). Los primeros datos obvios se reflejan claramente en la combinación de datos georreferenciados y clustering, y se observa como en horarios matinales, las personas se desplazan desde barrios periféricos a lugares de concentración comercial o industriales, y por las tardes este proceso se revierte.

\section{Conclusiones}
\label{sec:conclusion}
Este trabajo de análisis de datos permitió al cliente contar con herramientas para conocer de forma profunda y con altísimo nivel de detalle la distribución de la demanda. Esto permite agregar valor a la empresa mediante varios mecanismos, ya que conocer el detalle de la demanda habilita el uso de herramientas comerciales en forma sistemática e informada, que de otra forma se aproximan por la intuición o la experiencia. La intuición no siempre coincide con la situación real y actual en la dinámica del negocio, ya que refleja el conocimiento de muchos años y una visión en algunos casos subjetiva de una realidad cambiante. Para poder mejorar hay que saber medir. Big data nos posibilita medir en tiempo real y con alta definición.
%
%

\clearpage
\addtocmark[2]{Author Index} 
\renewcommand{\indexname}{Author Index}
\clearpage
\addtocmark[2]{Subject Index} 
\markboth{Subject Index}{Subject Index}
\renewcommand{\indexname}{Subject Index}
\end{document}